\documentclass[11pt,a4paper]{article}
\usepackage{cite,indentfirst}
\usepackage{amssymb,amsmath}
\usepackage[dvips]{epsfig}
\usepackage{epsf}
\usepackage{graphics}

\def\mprp{\mbox{\tiny $\bot$}}
\def\mprl{\mbox{\tiny $\|$}}

\def\beq{\begin{eqnarray}}
\def\eeq{\end{eqnarray}}
\def\ee{\varepsilon}
\def\lm{\lambda}
\newcommand{\prp}[1]{#1_{\mbox{\tiny $\bot$}}}
\newcommand{\prl}[1]{#1_{\mbox{\tiny $\|$}}}

\def\1{1 \to 1 \, 2}
\def\2{1 \to 2 \, 2}

\def\P{{\cal P}}

\title{Electromagnetic Processes In Strongly Magnetized Plasma}

\author{M. V. Chistyakov, D. A. Rumyantsev \\[3mm]
{\small\it Division of Theoretical Physics,} \\
{\small\it Yaroslavl State (P.G.~Demidov) University,} \\
{\small\it Sovietskaya 14, 150000 Yaroslavl, Russian Federation}\\
{\small\tt E-mail: mch@uniyar.ac.ru, rda@uniyar.ac.ru} }

\date{}

\begin{document}

\maketitle

\thispagestyle{empty}

\begin{abstract}

The electromagnetic processes of Compton scattering and photon
splitting/merging are investigated in the presence of strongly
magnetized electron-positron plasma. The influence of these
processes on the radiation transfer in the astrophysical environment
is studied. In particular, the contribution of the processes under
consideration in coefficients of the transfer equation is
calculated. We show the importance of photon splitting/merging
contribution and  taking into account of photon dispersion and wave
function renormalization in strong magnetic field and plasma.
\end{abstract}

\section{Introduction}

\begin{figure}[htb]
\centerline{\includegraphics[scale=0.5]{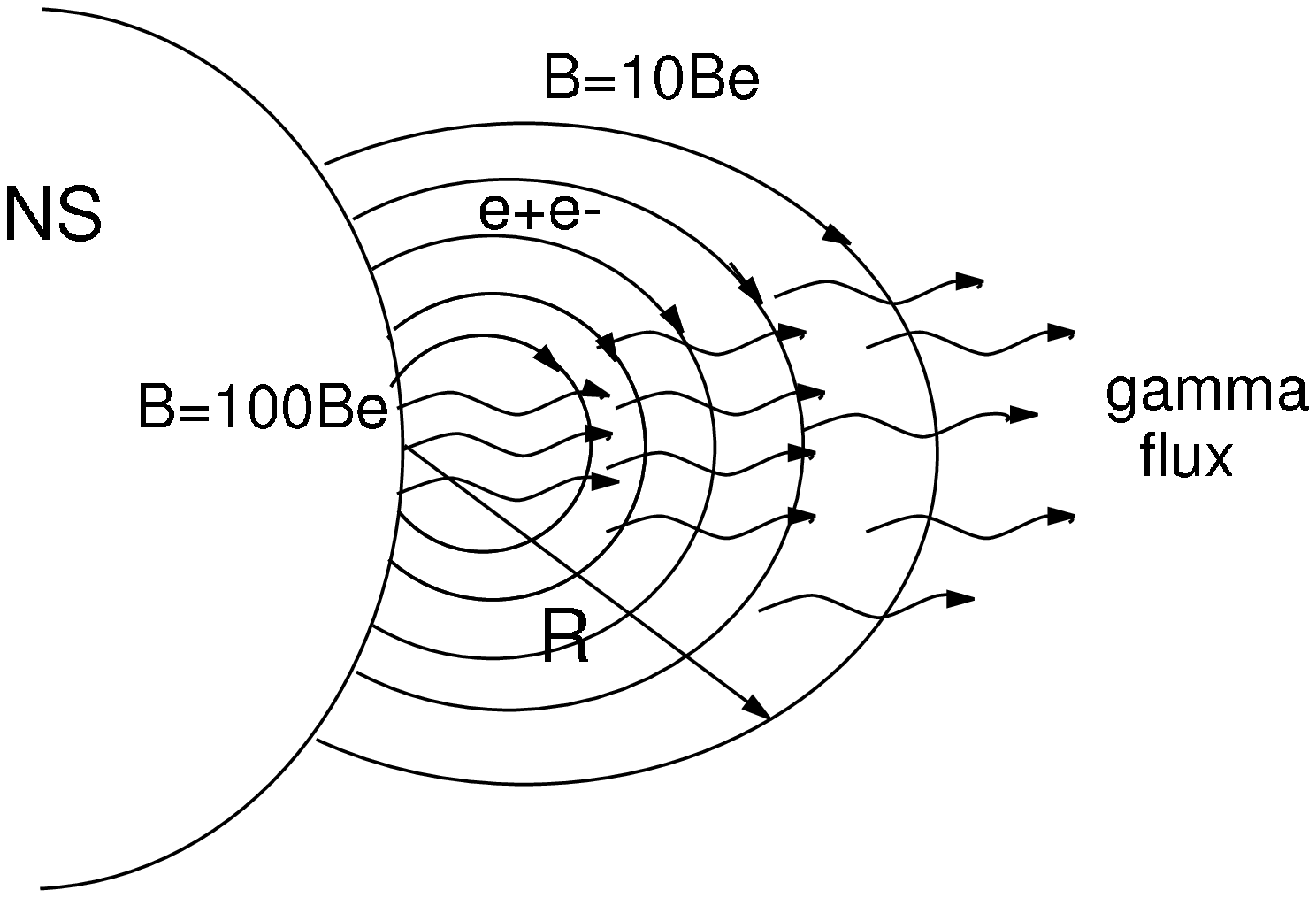}}
\caption{Radiation transfer in Tompson \& Duncan model of SGR
burst~\cite{Duncan:1995}.} \label{fig:geom}
\end{figure}

Magnetars are extremely interesting objects both from the physical
and astrophysical point of view. From one hand they are associated
with SGR and AXP pulsars considered as isolated neutron stars with
unusual spectral properties. From the other hand they allow one to
investigate different phenomena taking place in super strong
magnetic field condition not available elsewhere. Magnetic field
strength of magnetar is believed to be  $B \sim 10^{14}-10^{16}$
G~\cite{Duncan:1992,Duncan:1995,Duncan:1996}, i.e. $B \gg B_e$,
where $B_e = m^2/e \simeq 4.41\times 10^{13}$~G~\footnote{We use
natural units $c = \hbar = k = 1$, $m$ is the electron mass,  $e >
0$ is the elementary charge.} is the critical magnetic field. The
spectra analysis of these objects is also providing evidence for the
presence of electron-positron plasma in magnetar environment. It is
well-known that strong magnetic field and/or plasma could influence
essentially on different quantum
processes~\cite{Duncan:2000pj,Lai:2001,KM_Book,Harding:2006qn}. One
of such phenomena is the radiation transfer in strongly magnetized
plasma. This process is connected with the SGR and AXP spectral
formation. Moreover it is the crucial ingredient of the models of
SGR burst (see e.g.~\cite{Duncan:1995}) where the creation of
magnetically trapped high temperature ($\sim 1$  MeV) plasma
fireball is assumed (see Fig.~\ref{fig:geom}). It also defines the
cooling rate of the outer crust of magnetar~\cite{Yakovlev2000}.

The various studies indicate that electromagnetic processes such as
Compton scattering and photon splitting $\gamma \to \gamma \gamma$
(merging $\gamma \gamma \to \gamma$) could play a crucial role in
these models.

In the present work the influence of these processes on radiation
transfer is investigated in the presence of strong magnetic field
and electron positron plasma, when the magnetic field strength $B$
is the maximal physical parameter, namely $\sqrt{eB} \gg T,\, \mu \,
\omega, \, E$. Here $T$ is the plasma temperature, $\mu$ is the
chemical potential, $\omega$ and $E$ is the initial photon and
electron (or positron) energies. In this case almost all electrons
and positrons in plasma are on the ground Landau level.

The main goal of this talk is to demonstrate that the
self-consistent accounting of strong magnetic field and dense plasma
influence is necessary for the correct description of radiation
transfer.

\section{Photon dispersion properties}

The propagation of the electromagnetic radiation in any active
medium is convenient to describe in terms of normal modes
(eigenmodes). In turn, the polarization and dispersion properties of
normal modes are connected with eigenvectors and eigenvalues of
polarization operator correspondingly. In the case of strongly
magnetized plasma
 in the one loop approximation
the eigenvalues of the
polarization operator can be derived from the previously obtained
results~\cite{Rojas1979,Rojas1982,Shabad}:
\begin{eqnarray}
\P^{(1)}(q) &\simeq& - \frac{\alpha}{6 \pi}\,
\left [ q_{\mbox{\tiny $\bot$}}^2 +
\sqrt{q_{\mbox{\tiny $\bot$}}^4 +
\frac{(6N\omega)^2 q^2}{q_{\mprl}^2}}\, \right ] - q^2\, \Lambda(B) , \label{P1}
\\[2mm]
{\cal P}^{(2)}(q) &\simeq& -\frac{2 eB \alpha}{\pi}\left[
H\left(\frac{q^2_{\mprl}}{4m^2} \right) + {\cal J}(\prl{q}) \right]
- q^2 \, \Lambda(B), \label{P2}
\\[2mm]
\P^{(3)}(q) &\simeq&  - \frac{\alpha}{6 \pi}\,
\left [ q_{\mbox{\tiny $\bot$}}^2  -\sqrt{q_{\mbox{\tiny $\bot$}}^4 +
\frac{(6N\omega)^2 q^2}{q_{\mprl}^2}}\, \right ] - q^2\, \Lambda(B) ,
\label{P3}
\end{eqnarray}

where
$$
\Lambda(B) = \frac{\alpha}{3 \pi}\,\left[1.792 - \ln
(B/B_e)\right], \quad
N = \int \limits_{-\infty}^{+\infty} dp_z \,
\left [f_{-}(E) - f_{+}(E)\right ],
$$
\beq
\nonumber
{\cal J}(\prl{q}) = 2 \prl{q}^2 m^2 \int \frac{dp_z}{E} \,
\frac{f_{-}(E) + f_{+}(E)}
{(\prl{q}^2)^2 - 4 \prl{(pq)}^2}\,, \qquad E = \sqrt{p_z^2 + m^2},
\eeq
$f_{\pm}(E) \, = \, [e^{(E \, \pm \, \mu)/T} \, + \, 1]^{-1}$
 are the electron (positron) distribution functions,
\beq
\label{eq:H0}
&&H(z)=\frac{1}{\sqrt{z(1 - z)}} \arctan \sqrt{\frac{z}{1 - z}} - 1,
\quad 0 \leqslant z \leqslant 1,
\\ [3mm]
\label{eq:H1} &&H(z) = - \frac{1}{2\sqrt{z(z-1)}} \ln
\frac{\sqrt{z} + \sqrt{z-1}}{\sqrt{z} - \sqrt{z-1}}  - 1 +
\,\frac{i\pi}{2\sqrt{z(z-1)}}, \quad z > 1. \eeq

\noindent Here $z = q_{\mprl}^2/(4m^2)$, the four-vectors with
indices $\bot$ and $\parallel$ belong to the Euclidean \{1,
2\}-subspace and the Minkowski \{0, 3\}-subspace correspondingly in
the frame were the magnetic field is directed along $z$ (third)
axis; $(ab)_{\mprp} = (a \Lambda b) = a_\alpha \Lambda_{\alpha
\beta} b_\beta$, $(ab)_{\mprl} = (a \tilde \Lambda b) = a_\alpha
\tilde \Lambda_{\alpha \beta} b_\beta$, where the tensors
$\Lambda_{\alpha \beta} = (\varphi \varphi)_{\alpha \beta}$,\,
$\widetilde \Lambda_{\alpha \beta} = (\tilde \varphi \tilde
\varphi)_{\alpha \beta}$, with equation $\widetilde \Lambda_{\alpha
\beta} - \Lambda_{\alpha \beta} = g_{\alpha \beta} = diag(1, -1, -1,
-1)$ are introduced. $\varphi_{\alpha \beta} = F_{\alpha \beta} /B$
and ${\tilde \varphi}_{\alpha \beta} = \frac{1}{2}
\varepsilon_{\alpha \beta \mu \nu} \varphi_{\mu \nu}$ are the
dimensionless field tensor and dual field tensor correspondingly.

\begin{figure}[htb]
\centerline{\includegraphics[scale=0.8]{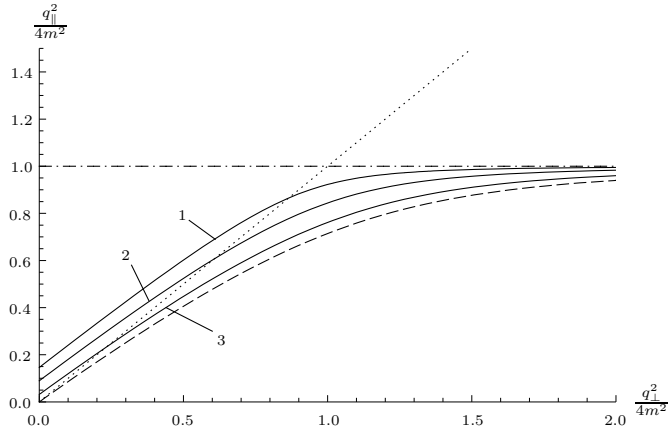}}  \caption{Photon
dispersion laws in strong magnetic field $B/B_e = 200$ and neutral
plasma vs. temperature:  $T = 1$ MeV -- 1,
 $T = 0.5$ MeV -- 2 and $T = 0.25$ MeV -- 3.
Photon dispersion without plasma is depicted by dashed line. Dotted
line corresponds to the vacuum dispersion law, $q^2 = 0$. The angle
between the photon momentum and the magnetic field direction is
$\pi/2$. \label{fig:dis1}}
\end{figure}

The dispersion properties of the normal modes could be defined from the
dispersion equations
\beq
q^2 - \P^{(\lm)}(q) = 0 \qquad (\lm = 1, 2, 3).
\label{disper}
\eeq
Their analysis shows that  1 and 2  modes with polarization vectors
\beq
\ee_\alpha^{(1)}(q) = \frac{(q \varphi)_\alpha}{\sqrt{q_{\mprp}^2}},
\qquad
\ee_\alpha^{(2)}(q) = \frac{(q \tilde \varphi)_\alpha}
{\sqrt{q_{\mprl}^2}}.
\label{epsilon}
\eeq
are only physical ones in the case under consideration, just as it is in the
pure magnetic field~\footnote{ Symbols 1 and 2 correspond to
the $\|$ and $\perp$ polarizations in pure magnetic
field~\cite{Adler:1971} and $E$- and $O$- modes in magnetized
plasma~\cite{Duncan:1995}.}. However, it should be emphasized  that this
coincidence is
approximate to within $O(1/\beta)$ and $O(\alpha^2)$ accuracy.
\begin{figure}[htb]
\centerline{\includegraphics[scale=0.8]{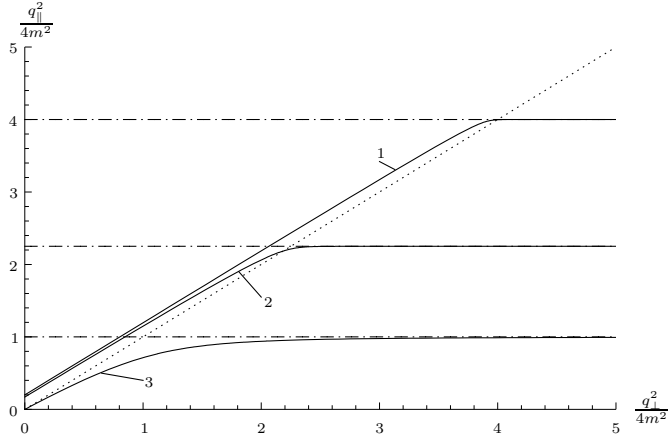}} \caption{Photon
dispersion in a strong magnetic field ($B/B_e = 200$) and degenerate
plasma vs. chemical potential $\mu = 1$ MeV -- 1, $\mu = 0.75$ MeV
-- 2 and without plasma -- 3. Dotted line corresponds to the vacuum
dispersion law, $q^2 = 0$. The angle between the photon momentum and
the magnetic field direction is $\pi/2$. \label{fig:dis0}}
\end{figure}

Notice, that in plasma only the eigenvalue $\P^{(2)}(q)$ is modified in
comparison with pure magnetic field case. It means that  the dispersion
law of the mode 1 is the same one as in the magnetized vacuum, where its
deviation from the vacuum law, $q^2 = 0$, is negligibly small.
From the other hand, the dispersion properties  of the mode 2 essentially differ from the
magnetized vacuum ones. In the Fig.~\ref{fig:dis1} --~\ref{fig:dis0}
  the photon dispersion in both
strong magnetic field and magnetized plasma are depicted at various
temperatures (for the charge-symmetric plasma) and chemical
potential (for the degenerate plasma). One can see that in the
presence of the magnetized plasma there exist the kinematical
region, where $q^2 >0$ contrary to the case of  pure magnetic field.
It is connected with the appearance  of the plasma frequency  in the
present of the real electrons and positrons which can be defined
from equation
\begin{equation}
\omega_{pl}^2 - \P^{(2)}(\omega_{pl}, {\mathbf k} \to 0 ) = 0.
\label{eq:omegapl}
\end{equation}
This fact could lead to the modification of the kinematics of the
different processes with photons. The analysis shows that the main
channels  of photon scattering and photon splitting/merging are

\begin{itemize}

\item{mode 1 (extraordinary photon):}
\begin{eqnarray}
&&\gamma_1 e^\pm \to \gamma_1 e^\pm, \gamma_1 e^\pm \to
\gamma_2e^\pm,
\gamma_1 \to \gamma_1 \gamma_2,
\nonumber \\[-1mm]
&&\gamma_1 \to \gamma_2 \gamma_2, \gamma_1 \gamma_2 \to \gamma_1,
\gamma_1 \gamma_1 \to \gamma_2;
\nonumber
\end{eqnarray}

\item{mode 2 (ordinary photon):}
\begin{eqnarray}
&&\gamma_2 e^\pm \to \gamma_2 e^\pm, \gamma_2 e^\pm \to \gamma_1
e^\pm, \gamma_2 \to \gamma_1 \gamma_1,
\nonumber\\[-1mm]
&&\gamma_2 \gamma_2 \to \gamma_1, \gamma_2 \gamma_1 \to
\gamma_1.
\nonumber
\end{eqnarray}

\end{itemize}

It follows from Eq.~(\ref{P2}) that the eigenvalue of the
polarization operator ${\cal P}^{(2)}$ becomes large near the electron-positron
pair production threshold.
This suggests that
the renormalization of the wave function for a photon
of this polarization should be taken into account:
\beq
\ee_{\alpha}^{(2)}(q) \to \ee_{\alpha}^{(2)}(q) \sqrt{Z_2}, \quad
Z^{-1}_2 = 1 - \frac{\partial {\cal P}^{(2)}(q)}{\partial \omega^2}.
\label{eq:eps}
\eeq

\section{Transfer equation}

In general case the propagation of photon modes through a magnetized
plasma  can be  described by the  following equations:
\begin{eqnarray} \frac{1}{r^2}\frac{d}{dr}\left (r^2D_1
\frac{dn_1}{dr}\right) + K_1 (\bar n - n_1) + S_{12} (n_2 - n_1) =
0,
\label{eq:TE1}\\
\frac{1}{r^2}\frac{d}{dr}\left (r^2D_2
\frac{dn_2}{dr}\right) + K_2 (\bar n - n_2) + S_{21} (n_1 - n_2) =
0,
\label{eq:TE2}
\end{eqnarray}

$$\bar n = \frac{\omega^3}{2\pi^2}\,f_{\omega}, \qquad f_{\omega} \, = \,
[\exp{(\omega/T)} \, - \, 1]^{-1}.$$
where $n_1, n_2$ are photon occupation numbers for extraordinary and
odinary modes, $f_\omega$ is photon distribution function,
$D_{\lambda}, K_{\lambda}, S_{\lambda \lambda'}$ are diffusion,
absorption and scattering coefficients for different photon modes
correspondingly ($\lambda = 1,2$) which can be  obtained by the
angle averaging of the photon splitting/merging  and photon
scattering rates:
\begin{eqnarray}
 D_\lambda &=& \int \frac{d\Omega}{4\pi} \ell_\lambda (\theta, r)
\cos^2{\theta}, \\
 K_\lambda &=& \int \frac{d\Omega}{4\pi} \left [W_{\lambda \to
\lambda^{'} \lambda^{''}} (\theta, r) + W_{\lambda \lambda^{'} \to
\lambda^{''}} (\theta, r) \right ],
\\
 S_{\lambda \lambda'} &=& \int \frac{d\Omega}{4\pi} W_{\lambda \to
\lambda'} (\theta, r),
\end{eqnarray}
where
\begin{eqnarray}
 \ell_\lambda = \left[\,\sum \limits_{\lambda'=1}^2 W_{\lambda \to
\lambda'}  + \sum \limits_{\lambda',\lambda''=1}^2 \left( W_{\lambda
\to \lambda^{'} \lambda^{''}} + W_{\lambda \lambda^{'} \to
\lambda^{''}} \right) \right]^{-1}. \label{eq:l}
\end{eqnarray}
In turn, the rates of the processes under consideration are given by
the following formulas:
\begin{eqnarray}
W_{\lambda  \to \lambda^{'}} &=& \frac{eB}{16 (2\pi)^4
\omega_{\lambda}} \int \mid {\cal M_{\lambda \lambda^{'}
}}\mid^2 Z_{\lambda}Z_{\lambda^{'}} \times
\label{eq:Wc}\\
&\times& f_{E}\, (1-f_{E'}) \, (1 + f_{\omega'}) \delta
(\omega_{\lambda}({\bf k}) + E - \omega_{\lambda^{'}}({\bf k'}) -
E') \frac{dp_z\,d^3 k^{'}}{ E E' \omega_{\lambda^{'}}};
\nonumber\\[2mm]
W_{\lambda \to \lambda^{'} \lambda^{''}} &=& \frac{1 -
(1/2)\delta_{\lambda' \lambda''}} {32 \pi^2 \omega} \int \mid {\cal
M_{\lambda \lambda^{'} \lambda^{''}}}\mid^2
Z_{\lambda}Z_{\lambda^{'}}Z_{\lambda^{''}} \times
\label{eq:Wsp} \\
&\times& (1 + f_{\omega'})(1 + f_{\omega''}) \delta
(\omega_{\lambda}({\bf k}) - \omega_{\lambda^{'}}({\bf k} - {\bf
k^{''}}) - \omega_{\lambda^{''}}({\bf k^{''}})) \frac{d^3
k^{''}}{\omega_{\lambda^{'}} \omega_{\lambda^{''}}},
\nonumber\\[2mm]
W_{\lambda \lambda^{'} \to \lambda^{''}} &=& \frac{1} {32 \pi^2
\omega} \int \mid {\cal M_{\lambda \lambda^{'} \lambda^{''}}}\mid^2
Z_{\lambda}Z_{\lambda^{'}}Z_{\lambda^{''}} \times
\label{eq:Wmer} \\
&\times& f_{\omega'}(1 + f_{\omega''}) \delta (\omega_{\lambda}({\bf
k}) + \omega_{\lambda^{'}}({\bf k'}) - \omega_{\lambda^{''}}({\bf k}
+ {\bf k^{'}})) \frac{d^3 k^{'}}{\omega_{\lambda^{'}}
\omega_{\lambda^{''}}}.
\nonumber
\end{eqnarray}
where $f_E$ is the electron distribution function, ${\cal
M}_{\lambda \lambda^{'}}$ and ${\cal M}_{\lambda \lambda^{'} \lambda
^{''}}$ are the partial amplitudes of the photon scattering and
photon splitting processes. Using the expressions for $Z_\lambda$
(\ref{eq:eps}) and taking account for photon dispersion properties
in energy conservation law inside $\delta$-function in
(\ref{eq:Wc})-(\ref{eq:Wmer}) one could obtained the self-consistent
result for the coefficients in transfer equations (\ref{eq:TE1}),
(\ref{eq:TE2}). To calculate the corresponding amplitudes in the
presence of strong magnetic field one should use the Dirac equation
solutions at the ground Landau level. For the electron propagator it
is relevant to use its asymptotic form~\cite{KM_Book}. It is
possible to present them in the covariant form. For Compton
scattering one has~\cite{RCh06}:
\begin{eqnarray}
{\cal M}_{1 1} &=& -\frac{8 \pi \alpha m}{eB}\,
\frac{(q \varphi q')(q \tilde \varphi q')}
{\sqrt{q^2_{\mprp} q'^2_{\mprp} (-Q^2_{\mprl})}},
\label{eq:M11}
\\[2mm]
{\cal M}_{1 2} &=& -\frac{8 \pi \alpha m}{eB}\,
\frac{(q \Lambda q')(q' \tilde \Lambda Q)}
{\sqrt{q^2_{\mprp} q'^2_{\mprl} (-Q^2_{\mprl})}},
\label{eq:M12}
\\[2mm]
{\cal M}_{2 1} &=& \frac{8 \pi \alpha m}{eB}\,
\frac{(q \Lambda q')(q \tilde \Lambda Q)}
{\sqrt{q^2_{\mprl} q'^2_{\mprp} (-Q^2_{\mprl})}},
\label{eq:M21}
\\
{\cal M}_{2 2} &=& 16 i \pi \alpha m\, \frac{\sqrt{q^2_{\mprl}
q'^2_{\mprl}}\,\sqrt{(-Q^2_{\mprl})}\, \varkappa} {(q \tilde \Lambda
q')^2 - \varkappa^2 (q \tilde \varphi q')^2}, \label{eq:M22}
\end{eqnarray}
\noindent where $\varkappa = \sqrt{1 - 4m^2/Q^2_{\mprl}}$ and
$Q^2_{\mprl} = (q - q')^2_{\mprl} <0$.
\begin{figure}[htb]
\centerline{\includegraphics[scale=0.9]{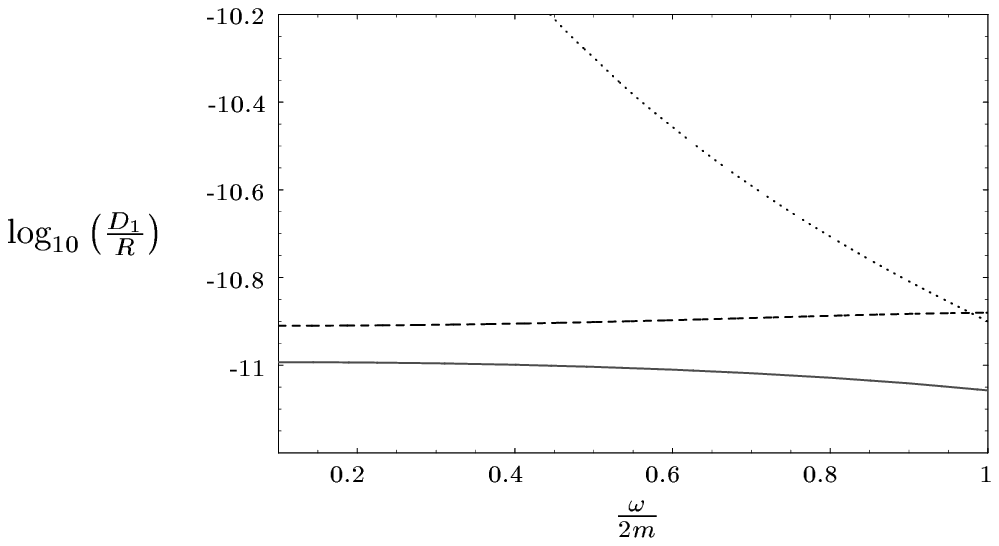}} \caption{
Diffusion coefficient for the 1-mode photon calculated at $B = 100
B_e, T = 1 MeV$ (solid line) and $B = 60 B_e, T = 0.5MeV$ (chain
line). Dotted line corresponds to the diffusion coefficient
calculated using approximation (\ref{eq:llow}).} \label{fig:D1}
\end{figure}

The amplitudes of photon splitting are given by the following
equations~\cite{RCh05}:
\begin{eqnarray}
{\cal M}_{1 1 2} &=& i 4\pi \left(
\frac{\alpha}{\pi}\right)^{\mbox{\tiny $3/2$}} \frac{(q^{'}\varphi
q^{''})(q^{'} \tilde \varphi q^{''})} {[\prp{q^{'2}} \prl{q^{''2}}
\prp{q^2}]^{\mbox{\tiny $1/2$}}} {\cal G} (q^{''}_{\mbox{\tiny
$\|$}}), \label{eq:M112}
\\[3mm]
{\cal M}_{1 2 2} &=& i 4\pi \left(
\frac{\alpha}{\pi}\right)^{\mbox{\tiny $3/2$}} \frac{(q'
q'')_{\mbox{\tiny $\|$}}} {[\prl{q'^2} \prl{q''^2}
q^2_{\mprp}]^{\mbox{\tiny $1/2$}}} \label{eq:M122}
\\[2mm]
&\times& \left \{(q q'')_{\mbox{\tiny $\bot$}} {\cal G}(q'_{\mbox{\tiny $\|$}}) +
(q q')_{\mbox{\tiny $\bot$}} {\cal G}(q''_{\mbox{\tiny $\|$}}) \right \},
\nonumber
\\[3mm]
{\cal M}_{2 1 1} &=& {\cal M}_{1 1 2} (q \leftrightarrow q''), \quad
\label{eq:M211}
\end{eqnarray}
where ${\cal G}(q) = H(\prl{q}^2/(4 m^2)) + {\cal J}(\prl{q})$.

The analysis of the  last equations  shows that all amplitudes much
smaller than ${\cal M}_{2 2}$. Therefore one could expect that mode
2 has the largest scattering absorption rate. It means that the
radiation transfer in the magnetically trapped plasma may be
described  as s diffusion of the 1-mode photons whereas 2-mode
photons are locked~\cite{Duncan:1995,Lyubarsky:2002}.
\begin{figure}[htb]
\centerline{\includegraphics[scale=0.9]{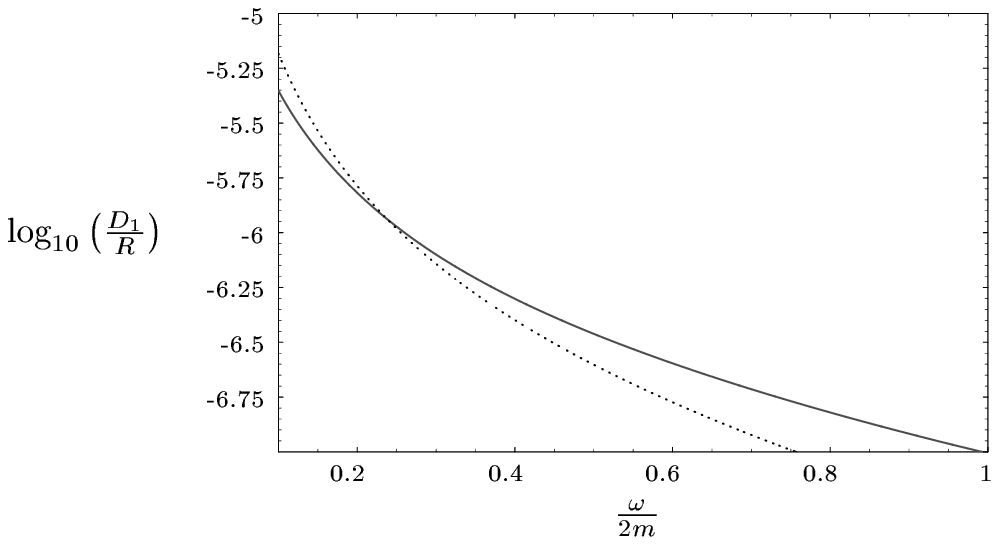}}
\caption{Diffusion coefficient for the 1-mode photon calculated at
$B = 10B_e,  T = 0.05 MeV$ (solid line). Dotted line corresponds ton
the diffusion coefficient calculated using approximation
(\ref{eq:llow}).}
 \label{fig:D1low}
\end{figure}

In general case, the calculation of the reactions rates
(\ref{eq:Wc})-(\ref{eq:Wmer}) is rather complicated mathematical
problem. However in some limiting cases it is possible to obtain the
simple expression for the rates. For example, in low temperature ($T
\ll m$) and low energy ($\omega \ll m$) limits the mean free path
(\ref{eq:l}) for the 1-mode photon in the charge symmetric plasma ($
\mu = 0$) can be presented in the following form:
\begin{eqnarray}
\ell_1^{-1} =  n_e \sigma_T (B_e \omega/Bm)^2 + (\alpha^3 \sin^6
{\theta}/2160\pi^2)(\omega/m)^5m, \label{eq:llow}
\end{eqnarray}
where $\sigma_T = \frac{8\pi}{3} \frac{\alpha} {m^2}$ is the
Thompson cross section and the number of electron (positron) density
in a strongly magnetized, charge-symmetric rarefied plasma can be
estimated as
\begin{eqnarray}
n_e \simeq eB \sqrt{\frac{mT}{2\pi^3}}\,e^{-m/T}. \label{eq:ne}
\end{eqnarray}
\begin{figure}[htb]
\centerline{\includegraphics[scale=0.8]{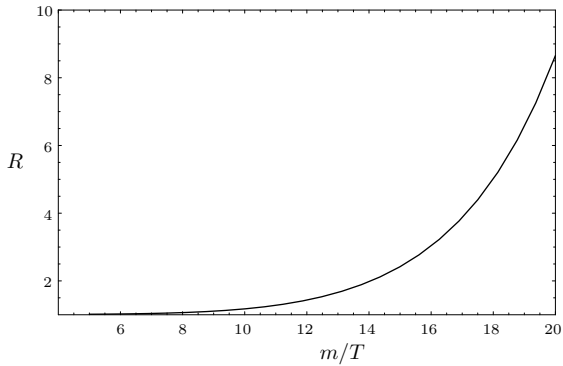}} \caption{The
ratio of the diffusion coefficients  without and with taking into
account of photon splitting process as a function of the inverse
temperature at $B = 200 B_e$.}
 \label{fig:R}
\end{figure}
In formula (\ref{eq:llow}) the fist term corresponds to the Compton
scattering process and the second one comes from the photon
splitting contribution. It is the estimation for the photon mean
free path that usually is used  in the radiation transfer analysis
in strongly magnetized plasma. Moreover, the process of photon
splitting/merging is not taking into account. We would like to show
that even in low energy limit this approximation is not appropriate.

\section{Discussion}

\begin{figure}[htb]
\centerline{\includegraphics[scale=0.9]{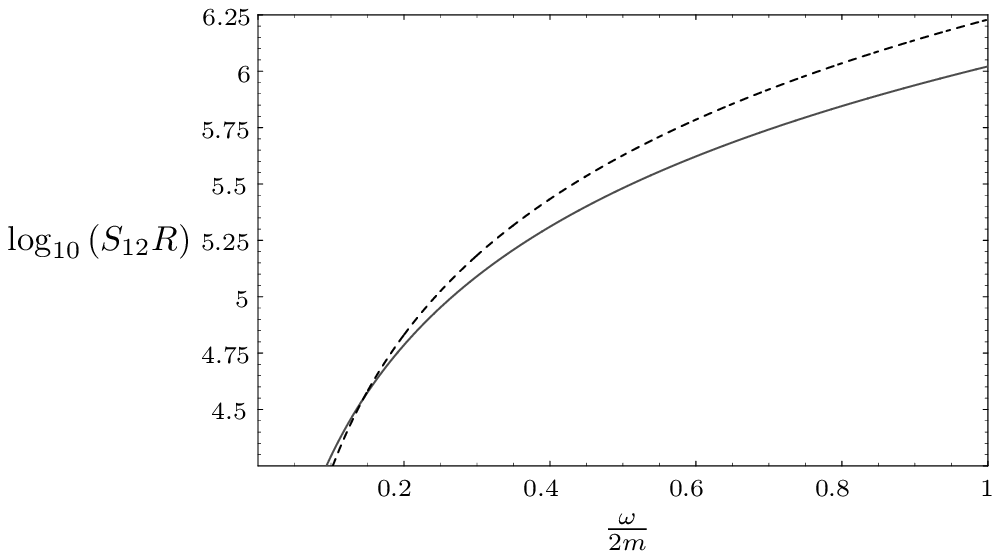}}
 \caption{Scattering
coefficient for the 1-mode photon calculated at $B = 10 B_e,  T =
0.05 MeV$ (solid line). Dashed line corresponds to the diffusion
coefficient calculated using approximation (\ref{eq:llow}).}
\label{fig:S}
\end{figure}

We have made the numerical calculation of the coefficients in
(\ref{eq:TE1}) and (\ref{eq:TE2}) in charge symmetric plasma. Our
results are represented in figures \ref{fig:D1}-\ref{fig:S}.

In figures \ref{fig:D1},  \ref{fig:D1low} and \ref{fig:S}  one can
see that diffusion coefficient calculated with taking account of
photon dispersion and large radiative correction strongly deviate
from the coefficient obtained by using approximation
(\ref{eq:llow}). In addition, in Fig. \ref{fig:R} the ratio of the
diffusion coefficient for only Compton scattering to  the diffusion
coefficient  with photon splitting is depicted. One can see that at
low temperatures the additional absorption process of photon
splitting leads to the significant decreasing of the diffusion
coefficient.

\begin{figure}[htb]
\centerline{\includegraphics[scale=0.8]{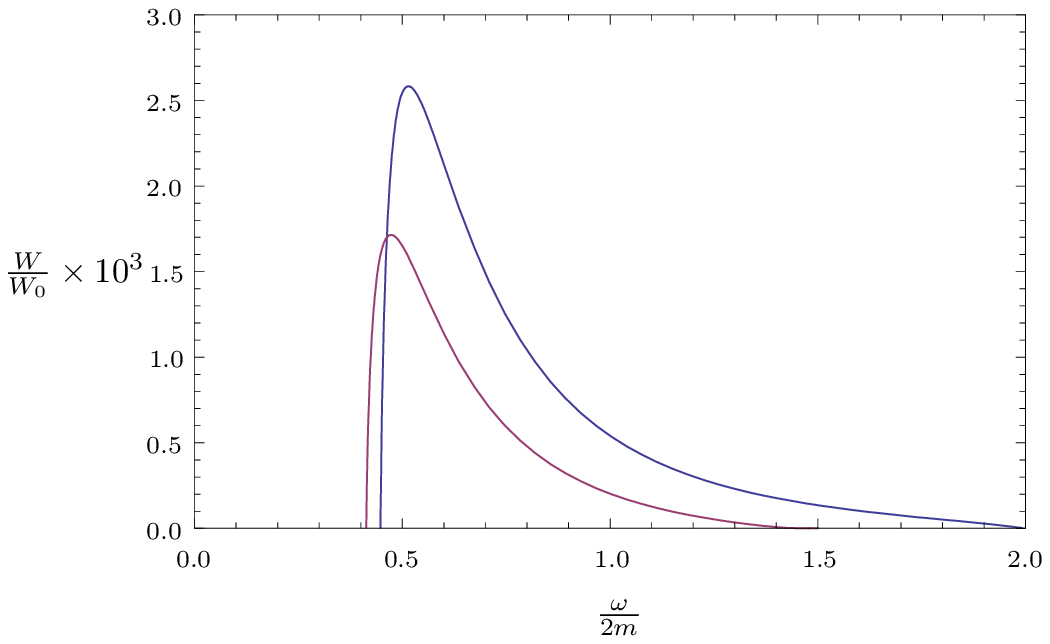}}
\caption{Absorption rate of the process  $\gamma_2 \to \gamma_1
\gamma_1$ at  $B = 200 B_e, \mu = 1.5 m$ (lower line), $\mu =  2m$
(lower line) , $W_0 \, = \, (\alpha/\pi)^3\, m \simeq 3.25\cdot 10^2
cm^{-1}$.} \label{fig:W211}
\end{figure}

\begin{figure}[htb]
\centerline{\includegraphics[scale=0.8]{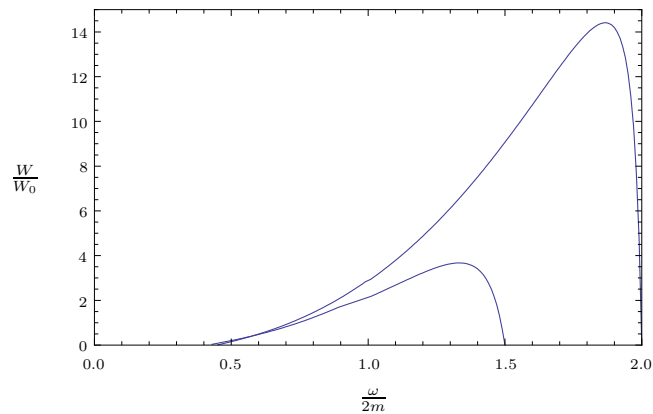}}
 \caption{Absorption rate of the process $\gamma_2 e^{\pm }\to \gamma_1 e^{\pm}$ at
 $B = 200B_e, \mu = 1.5 m$ (lower line), $\mu = 2m$ (upper line), $W_0 \, = \, (\alpha/\pi)^3\, m \simeq
3.25\cdot 10^2 cm^{-1}$.}\label{fig:W21}
\end{figure}

We have also analysed the problem of the radiation transfer in the
cold degenerate plasma. In this case the main channel of photon
splitting is $\gamma_2 \to \gamma_1 \gamma_1$. In the figures
\ref{fig:W211} and \ref{fig:W21} the contributions of photon
splitting and Compton scattering to photon mean free path are
depicted. One can see that in cold plasma the photon splitting
contribution is negligibly small in comparison with Compton
scattering. It means that the only process of photon scattering on
electrons defines the radiation transfer in cold plasma.

\section{Conclusion}

We have investigated the influence of strongly magnetized plasma on
the radiation transfer with taking into account of photon dispersion
and large radiative corrections.  We have studied the processes of
Compton scattering and photon splitting into two photons and
calculated their contribution in transfer equation coefficient. The
main conclusions of this work are

\begin{itemize}
\item
Photon dispersion and radiative correction in strongly magnetized
plasma could essentially influence on the radiation transfer
process.

\item
In charge symmetric plasma ($\mu = 0$) it is necessary to take into
account the processes of photon splitting and photon merging.

\item
In strongly degenerate plasma the influence of the photon splitting
and photon merging processes on radiation transfer is negligibly
small.
\end{itemize}

\section*{Acknowledgments}

We express our deep gratitude to the organizers of the Seminar
``Quarks-2008'' for warm hospitality. This work was supported in
part by the Russian Foundation for Basic Research under the Grant
No.~07-02-00285-a, and by the Council on Grants by the President of
the Russian Federation for the Support of Young Russian Scientists
and Leading Scientific Schools of Russian Federation under the Grant
No.~NSh-497.2008.2 and No. MK-732.2008.2 (MC).


\end{document}